\documentclass[pra,twocolumn,amsmath,amssymb,groupaddress,longbibliography]{revtex4-1}
\usepackage{graphicx}
\usepackage{amsthm}
\usepackage{amsfonts}
\usepackage{amssymb}
\usepackage{array}
\usepackage{amsmath}
\usepackage{verbatim} 
\usepackage{hyperref}
\usepackage{color}
\usepackage{bbold}
\usepackage{epstopdf}
\usepackage{mathtools}
\usepackage{bm,mathptmx,enumitem}
\usepackage[utf8]{inputenc}

\newcommand{\ket}[1]{\left|{#1}\right>}

\newcommand{\tr}[1]{\mathrm{tr}\!\left\{#1\right\}}

\newcommand{\ML}{\widehat{\rho}_\textsc{ml}}

\def\ket#1{\left|#1 \right>}

\begin{document}
\title{Adaptive compressive tomography: a numerical study}
\author{D.~Ahn}
\affiliation{Department of Physics and Astronomy, Seoul National University, 08826 Seoul, Korea}
\author{Y.~S.~Teo}
\email{ys\_teo@snu.ac.kr}
\affiliation{Department of Physics and Astronomy, Seoul National University, 08826 Seoul, Korea}
\author{H.~Jeong}
\affiliation{Department of Physics and Astronomy, Seoul National University, 08826 Seoul, Korea}
\author{D.~Koutn{\'y}}
\affiliation{Department of Optics, Palack\'{y}  University,
	17. listopadu 12, 77146 Olomouc, Czech Republic}
\author{J.~\v{R}eh\'{a}\v{c}ek}
\affiliation{Department of Optics, Palack\'{y}  University,
	17. listopadu 12, 77146 Olomouc, Czech Republic}
\author{Z.~Hradil}
\affiliation{Department of Optics, Palack\'{y}  University,
	17. listopadu 12, 77146 Olomouc, Czech Republic}
\author{G.~Leuchs}
\affiliation{Max-Planck-Institut f\"ur  die Physik des Lichts,
	Staudtstra\ss e 2, 91058 Erlangen, Germany}
\author{L.~L.~S{\'a}nchez-Soto}
\affiliation{Max-Planck-Institut f\"ur  die Physik des Lichts,
	Staudtstra\ss e 2, 91058 Erlangen, Germany}
\date{\today}

\begin{abstract}
	We perform several numerical studies for our recently published adaptive compressive tomography scheme~[D.~Ahn \emph{et al.} Phys.~Rev.~Lett.~{\bf122}, 100404 (2019)], which significantly reduces the number of measurement settings to unambiguously reconstruct any rank-deficient state without any \emph{a priori} knowledge besides its dimension. We show that both entangled and product bases chosen by our adaptive scheme perform comparably well with recently-known compressed-sensing element-probing measurements, and also beat random measurement bases for low-rank quantum states. We also numerically conjecture asymptotic scaling behaviors for this number as a function of the state rank for our adaptive schemes. These scaling formulas appear to be independent of the Hilbert space dimension. As a natural development, we establish a faster hybrid compressive scheme that first chooses random bases, and later adaptive bases as the scheme progresses. As an epilogue, we reiterate important elements of informational completeness for our adaptive scheme.
\end{abstract}
\maketitle

\section{Introduction}  

The main aim of quantum state tomography \cite{Chuang:2000fk,lnp:2004uq,Teo:2015qs} is to reconstruct the unknown true quantum state $\rho_\text{t}$ from data ${\mathbb{D}}$ obtained after a set of measurements (collectively represented by the map $\mathcal{M}[\rho_\text{t}]\mapsto\mathbb{D}$) are performed. Since quantum mechanics constrains $\rho_\text{t}$ to be a positive operator of unit trace, it is unambiguously specified by $d^2-1$ free parameters. 

To fully characterize $\rho_\text{t}$, one performs an informationally complete (IC) set of $O(d^2)$ measurement outcomes $\Pi_j\geq0$ that form a positive operator-valued measure (POVM) ($\sum_j\Pi_j=1$). The corresponding physical probabilities $p_{\text{t},j}$ relate the POVM and $\rho_\text{t}$ \emph{via} Born's Rule, expressed by either $p_{\text{t},j}=\mathrm{tr}\{\rho_\text{t}\,\Pi_j\}$ or $\mathcal{M}[\rho_\text{t}]=\bm{p}_\text{t}$. Realistically, $\bm{p}_\text{t}$ is always inaccessible. Rather, $\mathbb{D}$ is collected from $N<\infty$ independent sampling copies, so that one acquires the relative frequency $\nu_j=n_j/N$ for each outcome $\Pi_j$ of $n_j$ data counts. Working with $\mathbb{D}\equiv\{\nu_j\}$, the \emph{maximum-likelihood} (ML) method \cite{Rehacek:2007ml,Teo:2011me,Teo:2015qs,Shang:2017sf} may be used to supply the unique state estimator $\ML\geq0$ and its corresponding ML physical probabilities ($\bm{p}_\textsc{ml}$). In the limit $N\gg1$, we have $\ML\rightarrow\rho_\text{t}$. In both hypothetical noiseless and practical noisy situations, we say that operationally, \emph{$\mathbb{D}$ is IC if $\widehat{\rho}_\textsc{ml}$ is unique with respect to the whole state space}~\cite{Busch:1989aa,Prugovecki:1977aa}. Incidentally, this notion is synonymous to that of ``strictly-complete'' in \cite{Kalev:2015aa, Baldwin:2016cs}.

For complex quantum systems in high-dimensional states, measuring $O(d^2)$ outcomes quickly turns into a resource-intensive practical problem~\cite{Haffner:2005aa,Titchener:2018aa}. A prototypical strategy to circumvent this problem is to first assume \emph{a priori} that $\mathrm{rank} \{\rho_\text{t}\}\leq r$ for a given small $r \ll d$, and next invoke the conventional method of \emph{compressed sensing} (CS)~\cite{Donoho:2006cs,Candes:2006cs,Candes:2009cs} to find a unique rank-deficient state estimator~\cite{Gross:2010cs,Kalev:2015aa, Baldwin:2016cs,Steffens:2017cs,Riofrio:2017cs}. This widely-held standard, however, faces two practical issues. To start off, the \emph{a priori} assumption about $r$ requires additional verification before it can be applied to CS, since the resulting estimator accuracy hinges on the validity of this assumption. Next, as one has no means of validating the final estimator self-consistently, the usual solution is to compare the estimator with some assumed target state~\cite{Kalev:2015aa,Steffens:2017cs,Riofrio:2017cs}. In the presence of experimental errors, there is simply no guarantee whether such a comparison is actually trustworthy.

In Ref.~\cite{Ahn:2019aa}, we developed and experimentally carried out an adaptive compressive tomography (ACT) protocol that does not rely on any \emph{a priori} information about the quantum state apart from its dimension $d$. It consists of a self-consistent informational completeness certification (ICC) stage and an adaptive measurement optimization stage, both of which operate only on accumulated data~(we suggest the reference to Appendix~\ref{app:acro} as the list of acronyms lengthens). For practical implementation of ACT, we considered adaptive choices of von~Neumann (orthonormal) measurement bases that are generally more easily realized in the laboratory than arbitrary POVMs, as well as the product variant PACT that is compatible with many-body systems by utilizing local product bases instead of entangled ones as in ACT. 

In this article, we shall perform a series of numerical studies to explore (P)ACT. In particular (a)~we shall demonstrate, up to a reasonably large $d$ and $r$, that in terms of the number of IC bases $k_\textsc{ic}$ needed for unique reconstructions, our adaptive schemes outperform known random bases measurements. (b)~We next show that the $k_\textsc{ic}$ scaling behaviors of both ACT and PACT match the respective ones of the element-probing Baldwin--Flammia POVMs and Baldwin--Goyeneche bases reported in \cite{Baldwin:2016cs}, thereby effectively provides conjectured asymptotic scaling behaviors for both adaptive schemes in the limits $r\ll d$ and $d\gg2$. More specifically, we have $k_\textsc{ic}=2r+2$ for ACT and $k_\textsc{ic}=4r+1$ for PACT. Finally, (c)~we present a ``random-adaptive'' hybrid compressive tomography scheme that combines the respective speed and compressive advantages of random and adaptive strategies.

The article is organized as follows. Sections~\ref{sec:CS} and \ref{sec:ACT} respectively provide a preliminary introduction to CS and ACT for subsequent discussions. After providing some details about the simulation specifications in Sec.~\ref{sec:sim}, all numerical results are then presented in Sec.~\ref{sec:num_res}. We end the article with an additional remarks on the concept of ``IC'' in ACT in Sec.~\ref{sec:epilogue}.

\section{Compressed-sensing tomography}
\label{sec:CS}

The concept of CS~\cite{Donoho:2006cs,Candes:2006cs,Candes:2009cs} provides one approach to recover an unknown signal \emph{of known degree of sparsity} by performing a small set of specialized compressive measurement outcomes, and has since been widely adopted in signal processing \cite{Rani:2018aa}. In the context of quantum state tomography~\cite{Gross:2010cs,Kalev:2015aa, Baldwin:2016cs,Steffens:2017cs,Riofrio:2017cs}, the low-rank assumption often taken for granted, that is $\mathrm{rank}\{\rho_\text{t}\}\leq r$ for a \emph{known} and sufficiently small integer $r$, permits the utilization of CS to uniquely reconstruct $\rho_\text{t}$ with very few outcomes. While the primary focus began with random Pauli observables $M=O(rd(\log\,d)^2)$~\cite{ Gross:2010cs,Flammia:2012aa}, compressive measurements of more variety were later constructed~\cite{Baldwin:2016cs}.
  
\begin{figure}[t] 
 	\centering
    \includegraphics[width=0.9\columnwidth]{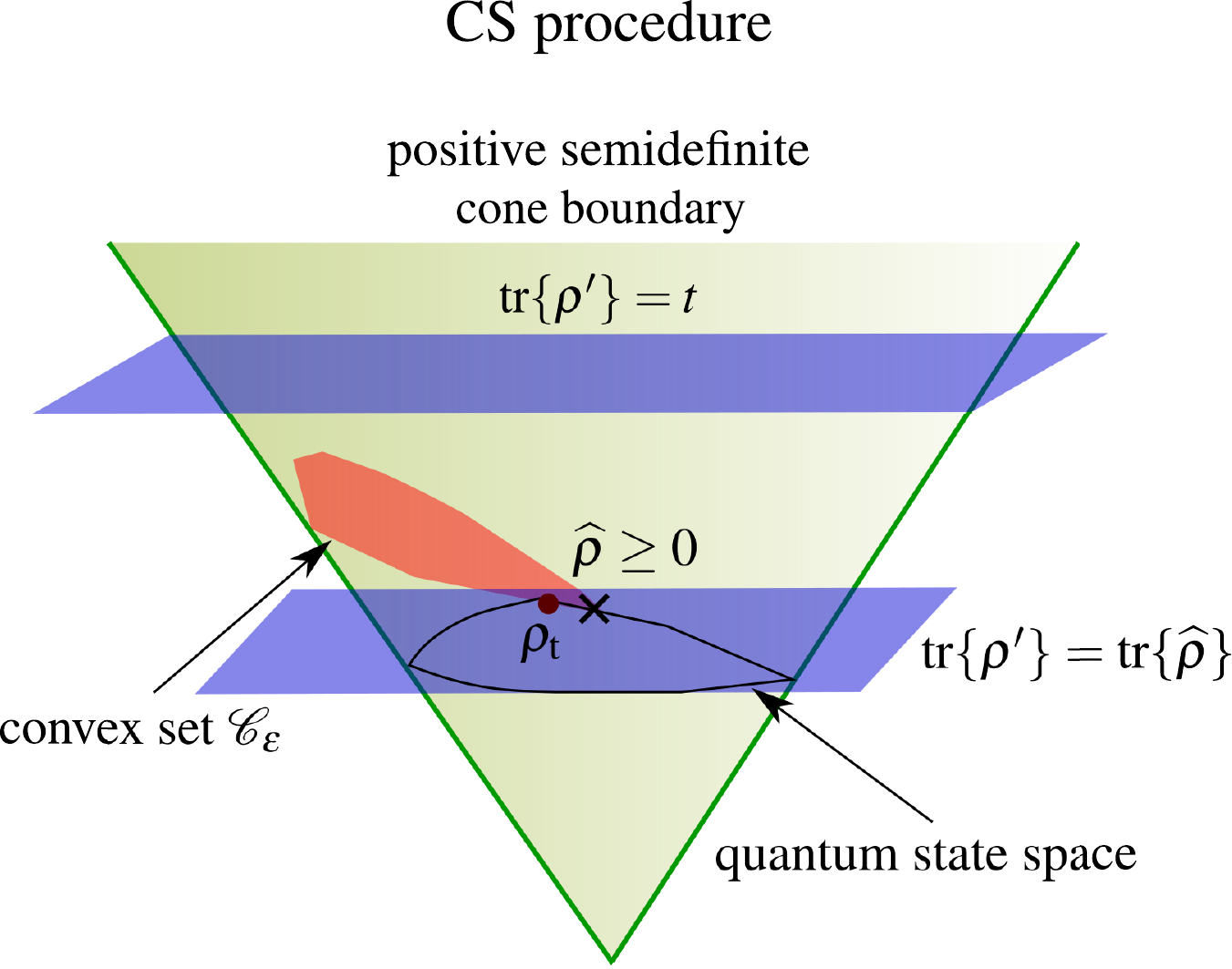}
  	\caption{\label{fig:CS}Pictorial review of CS. A convex set $\mathcal{C}_\epsilon$ is formed from quantum states that satisfy the inequality $\|\mathcal{M}[\rho']-\bm{\nu}\|\leq\epsilon$. For an appropriate compressive measurement map $\mathcal{M}$, trace minimization over the state space yields a unique state estimator $\widehat{\rho}\geq0$ that is typically near $\rho_\text{t}$ for $N\gg1$.}
\end{figure}
  
\subsection{Summary of the compressed-sensing procedure}  
Based on the assumed prior information, for measurement map $\mathcal{M}$ to be compressive and IC with respect to $\rho_\text{t}$, it is sufficient to satisfy the so-called restricted isometry property~\cite{Candes:2005aa,Candes:2006aa}. Without spelling out unnecessary mathematical details, we note that there also exist other kinds of compressive measurements without such a property~\cite{Baldwin:2016cs}. With a compressive $\mathcal{M}$, one seeks the unique estimator $\widehat{\rho}$ from the data convex set $\mathcal{C}$, which contains all possible states that satisfies the constraint $\mathcal{M}[\widehat{\rho}]=\bm{p}_\text{t}$ in the noiseless regime. 

One possible way is to choose $\widehat{\rho}$ that has the lowest rank out of $\mathcal{C}$. It can be shown that for $\mathcal{M}$ satisfying the restricted isometry property above certain threshold degree of orthonormality, this nonconvex optimization is equivalent to the convex optimization of the trace-norm (nuclear-norm) minimization~\cite{Recht:2010aa}, where the rank and the trace-norm are respectively the operator version of the $l_0$-norm and the $l_1$-norm meant for signal recovery~\cite{Candes:2006cs}. This ensures that the set of states consistent with $\mathbb{D}$ intersects the state space at exactly one lowest-rank state that coincides with $\rho_\text{t}$ if $N\gg1$. With real dataset $\mathbb{D}$ and positivity~(see also Fig.~\ref{fig:CS}), the CS algorithm (proven to work with measurements possessing restricted isometry property) that uniquely identifies a state estimator may be simplified~\cite{Kalev:2015aa} into 

\begin{center}
	\begin{minipage}[c][5cm][c]{0.9\columnwidth}
		\noindent
		\rule{\columnwidth}{1.5pt}\\
		\textbf{CS procedure with positivity}\\[1ex]
		For known $r$ and data $\mathbb{D}=\{\nu_j\}\equiv\bm{\nu}$:
		\begin{enumerate}
		\item Minimize $\tr{\rho'}$, subject to
		\begin{itemize}
			\item $\|\mathcal{M}[\rho']-\bm{\nu}\|\leq\epsilon$ for some $\epsilon>0$ that depends on noise, 
			\item $\rho'\geq0$.
		\end{itemize}
		\item Trace-renormalize the optimal $\rho'$.\\[-5ex]
		\end{enumerate}	
		\rule{\columnwidth}{1.5pt}
	\end{minipage}
\end{center}
\noindent
It can be shown that the above optimization routine leads to perfect recovery of $\rho_\text{t}$ for noiseless data ($\bm{\nu}\rightarrow\bm{p}_\text{t}$ with $\epsilon\rightarrow0$).

\subsection{Typical compressive measurements}
\label{subsec:csmeas}

For subsequent numerical comparisons, we shall consider a few well-known compressive POVMs that were applied to quantum state tomography. The first classic compressive measurement is the set of random Pauli bases~(RP) for $n$-qubit systems~\cite{Gross:2010cs,Kalev:2015aa}, which maximally comprise $3^n$ sets of $n$ tensor products of the standard qubit Pauli operators $\sigma_x$, $\sigma_y$ and $\sigma_z$.

On the other hand, the independent studies of two other types of measurements~\cite{Flammia:2005aa,Goyeneche:2015aa} related to pure-state distinction eventually led to the construction of rank-$r$ generalized compressive measurements~\cite{Baldwin:2016cs}. These are the element-probing Baldwin--Flammia~(BF) POVM and Baldwin--Goyeneche~(BG) bases that are constructed using the mathematical concepts of Schur complement and block-diagonalization of the density matrix. More importantly, these compressive measurements have analytical performance scaling behaviors in the regime $r\ll d$. Upon denoting the number of IC outcomes as $M_\textsc{ic}$, the BF POVM gives $M_\textsc{ic}=(2d-r)r + 1$ [two outcomes more than the number of free parameters $(2d-r)r - 1$ for a rank-$r$ state], and the BG bases give $k_\textsc{ic}=4r+1$, or $M_\textsc{ic}=(4r+1)d$.

Random measurements first fueled the progress of CS~\cite{Candes:2006rm}. For state tomography, we look into two classes of random measurements. The first class is the set of random bases generated by Haar unitary operators~(RH) that are widely used in quantum information theory~\cite{Mezzadri:2007qr,Cwiklinski:2013aa,Russell:2017aa,Banchi:2018aa}. The second class is the set of eigenbases of random full-rank states~(RS) distributed uniformly according to the Hilbert-Schmidt measure~\cite{Zyczkowski:2003hs}. For more details regarding the numerical constructions of random bases, we refer the interested Reader to the Appendix~\ref{app:rand}. Finally, there exists yet another benchmark by Kech and Wolf~(KW) for arbitrary von Neumann bases, which states that  $k\geq4r\lceil\frac{d-r}{d-1}\rceil$ is sufficient to uniquely reconstruct any rank-$r$ state~\cite{Kech:2016aa}.

\subsection{Issues in conventional compressed sensing}

While the CS procedure is a promising candidate for low-rank state recovery. There are two primary concerns that need to be addressed.

First, the proper set of compressive measurements is on the premise that the upper bound $r$ of the rank of $\rho_\text{t}$ is known. The assumption of $r$ must thus be experimentally justified. A categorical misclassification of $\rho_\text{t}$ would result in either a non-IC (too small an $r$) or an unnecessarily overcomplete dataset (too big an $r$). The former leads to an ambiguous set of estimators, while the latter overuses measurement resources.

Second, there exists no method in CS to systematically validate if the acquired $\mathbb{D}$ is truly IC in real experiments. Instead, the fidelity measure is commonly used as the indicator that the estimator is ``close'' to some prechosen target state. This approach, at the very least, requires yet another round of certification for these target states, and is evidently not the correct informational completeness indicator as it registers no information about the data convex set $\mathcal{C}$.

\section{Adaptive compressive tomography}
\label{sec:ACT}

Our recently proposed ACT~\cite{Ahn:2019aa} is able to deterministically compress IC datasets for any given rank-$r$ $\rho_\text{t}\equiv\rho_r$ without relying on any \emph{a priori} information or ad hoc assumptions about $\rho_r$ apart from its dimension $d$. The procedure of ACT is iterative and involves two main stages in every step. The first \emph{informational completeness certification} stage (ICC) decides, given the measured outcomes and accumulated data $\mathbb{D}$, whether the data convex set $\mathcal{C}$, which is again the set of states $\mathcal{C}=\{\rho'\}$ for which $\mathcal{M}[\rho']=\bm{p}_\textsc{ml}$ is singleton or not. The second \emph{adaptive strategy} provides protocols which adaptively chooses the next measurement at each step of ACT according to the accumulated data $\mathbb{D}$. The former would imply a unique state estimator consistent with an IC $\mathbb{D}$, and the procedure terminates. The latter would introduce appropriate protocols which significantly reduces the size of IC $\mathbb{D}$ compared to $O(d^2)$.

\begin{figure}[t] 
	\centering
	\includegraphics[width=1\columnwidth]{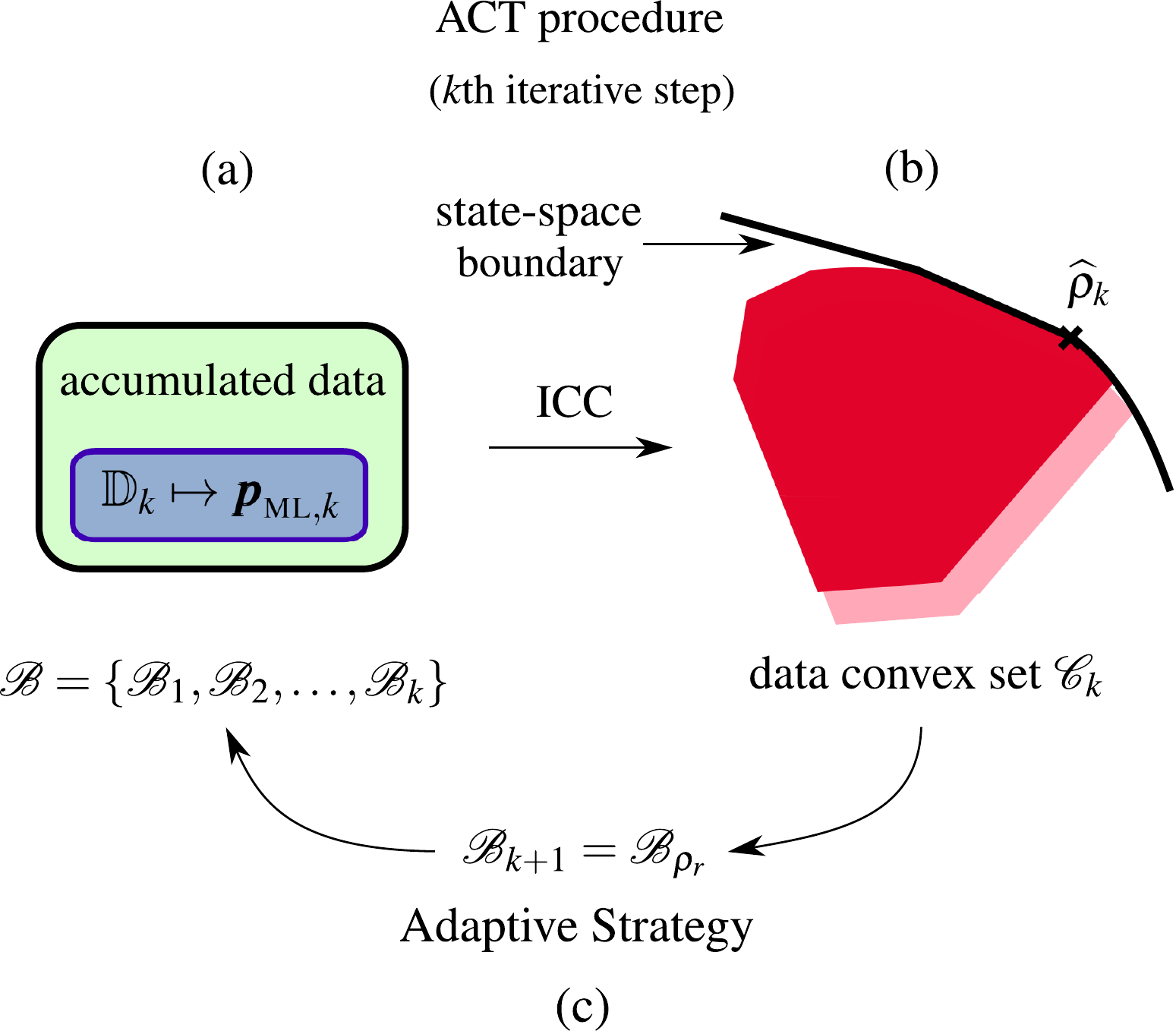}
	\caption{\label{fig:ACT}Pictorial review of ACT at the $k$th iterative step. (a)~The accumulated dataset $\mathbb{D}_k$ thus far undergoes ICC to verify if (b)~the data convex set $\mathcal{C}_k$ is a point or not. If not, (c)~ACT undergoes adaptive strategy that picks an optimal estimator $\widehat{\rho}_k$ and assigns its eigenbasis as the next von Neumann basis $\mathcal{B}_{k+1}$ to be measured. The cycle continues until $\mathcal{C}_{k=k_\textsc{ic}}$ becomes a point.}
\end{figure}

Throughout the article, we shall take the POVM to be sets of von Neumann bases that are each denoted by $\mathcal{B}$ and contains $d$ orthonormal projectors, which are the typical kind of measurement employed in laboratory. A particular iterative adaptation step of ACT is then to search for an optimal basis to measure in the next step (see Fig.~\ref{fig:ACT}).

\subsection{Informational completeness certification}

As ACT progresses iteratively and a sequence of von Neumann bases $\mathcal{B}=\{\mathcal{B}_1,\mathcal{B}_2,\ldots,\mathcal{B}_k\}$ are measured, the size $\zeta_k$ of the data convex set $\mathcal{C}_k$ at the $k$th iterative step of ACT indicates whether the corresponding dataset $\mathbb{D}_k$ of $k$ measured orthonormal bases is IC or not---$\mathbb{D}_k$ is IC if and only if $\zeta_k=0$. In this case, we denote the set of IC bases to be $k_\textsc{ic}=k$. In other words, the task of ICC is to verify if $\zeta=0$ (or close to some small numerical threshold value in practice).

\begin{figure}[t] 
	\centering
	\includegraphics[width=0.7\columnwidth]{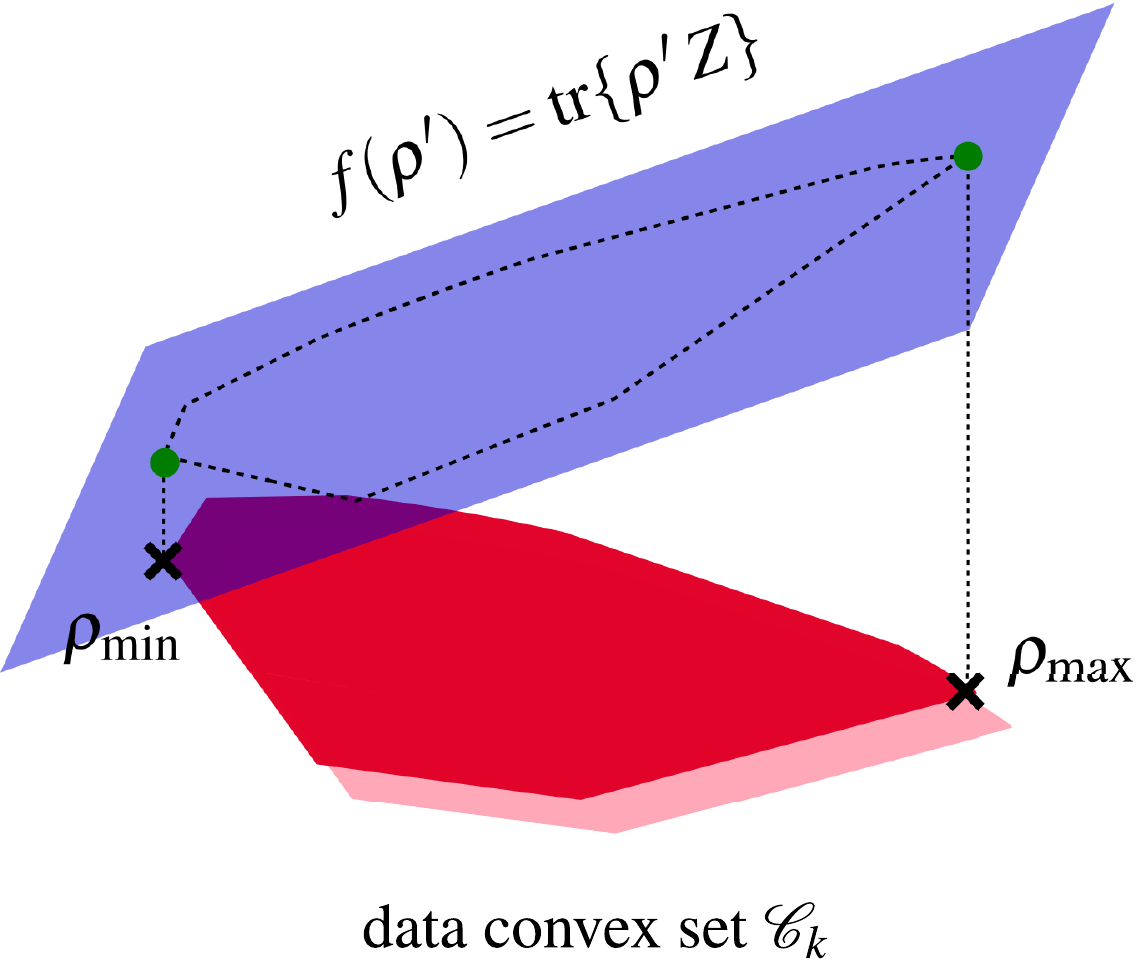}
	\caption{\label{fig:ICC}The ICC procedure. The linear function $f(\rho')=\tr{\rho' Z}$ (represented by a hyperplane) has global optimas at the edges of $\mathcal{C}_k$. The corresponding auxiliary extrema states $\rho_\text{max}$ and $\rho_\text{min}$ determine whether $\mathcal{C}_k$ is a singleton or not through $s_{\textsc{cvx},k}$.}
\end{figure}

Fortunately, whilst $\zeta_k$ is a complicated state-space integral with respect to some volume measure over $\mathcal{C}_k$, there is an alternatively much more feasible way to detect if $\zeta_k=0$. Using a randomly chosen full-rank state $Z$ throughout the entire run of ACT, if we define the linear function $f(\rho')=\mathrm{tr}\{\rho'\,Z\}$ for the variable state $\rho'$, then owing to the convexity of $\mathcal{C}_k$ \emph{regardless of whether data noise is present}, it straightforwardly follows that $\zeta_k=0$ iff $f_{\text{min},k}=f_{\text{max},k}$, where $f_{\text{min},k}$ and $f_{\text{max},k}$ are respectively the minimum and maximum values of $f$ in $\mathcal{C}_k$ (see Fig.~\ref{fig:ICC}). Hence, the validity of this if-and-only-if statement is robust against noise. This simple conclusion holds as long as $Z$ is neither $1/d$ nor possesses a kernel that contains $\rho'\in\mathcal{C}_k$, which are all measure-zero situations. Simply put, the business of ICC is a semidefinite program~\cite{Vandenberghe:1996sd} that evaluates the quantity $s_{\textsc{cvx},k}=(f_{\text{max},k}-f_{\text{min},k})/(f_{\text{max},1}-f_{\text{min},1})$ at every $k$th iterative step of ACT.

\begin{center}
	\begin{minipage}[c][6cm][c]{0.9\columnwidth}
		\noindent
		\rule{\columnwidth}{1.5pt}\\
		\textbf{ICC in the $k$th step}
		\begin{enumerate}
			\item Maximize and minimize $f_Z(\rho')=\tr{\rho' Z}$ for a fixed, randomly-chosen full-rank state $Z\neq1/d$ to obtain $f_{\text{max},k}$ and $f_{\text{min},k}$
			subject to
			\begin{itemize}
				\item $\rho'\geq0$\,, $\tr{\rho'}=1$\,,
				\item $\mathcal{M}[\rho']=\bm{p}_{\textsc{ml},k}$ for $\bm{p}_{\textsc{ml},k}$ obtained from the accumulated $\mathbb{D}_k$ thus far\,.
			\end{itemize}
			\item Compute $0\leq s_{\textsc{cvx},k}\leq1$ and check if it is smaller than some threshold $\epsilon$.
			\item If $s_{\textsc{cvx},k}<\epsilon$, terminate ACT. Continue otherwise.\\[-5ex]
		\end{enumerate}
		\rule{\columnwidth}{1.5pt}
	\end{minipage}
\end{center}

More specifically, $s_\textsc{cvx}$ is known as the \emph{size monotone} in the sense that with decreasing $\zeta_k$ or increasing $k$, $s_{\textsc{cvx},k}$ never increases in value for near-perfect $\mathbb{D}_k$. The reason for this is that with perfect $\mathbb{D}$s, we have the set hierarchy $\mathcal{C}_1\supseteq\mathcal{C}_2\supseteq\ldots\supseteq\mathcal{C}_{k_\textsc{ic}}$ as more linearly-independent bases are measured, so that  $f_{\text{max},k}$ monotonically decreases and $f_{\text{min},k}$ monotonically increases for a linear $f$, or $s_{\textsc{cvx},k+1}\leq s_{\textsc{cvx},k}$. Moreover, it is easy to see that if $f_{\text{max},k+1}-f_{\text{min},k+1}<f_{\text{max},k}-f_{\text{min},k}$, then $\mathcal{C}_{k+1}\subset\mathcal{C}_{k}$.

In terms of complexity, unlike the traditional NP-hard problem to determine whether a given set of POVM possesses the restricted isometry property~\cite{Bandeira:2013aa}, the complexity to decide if this POVM is IC with respect to $\mathbb{D}$ turns out to be, understandably, only as high as that for the semidefinite program employed in ICC. 
            
\subsection{Adaptive strategy}             
Given the collected dataset $\mathbb{D}_k$ and corresponding non-singleton data convex set $\mathcal{C}_k$ at the $k$th iterative step of ACT, one seeks the optimal basis to be measured in the $(k+1)$th step in order to converge ACT at a reasonably quick pace to the singleton $\mathcal{C}_{k=k_\textsc{ic}}$. 

If $\rho_r$ is a low-rank state ($r\ll d$), the natural motivation for a compressive adaptive strategy, given no other assumption, would be to actively seek the lowest-rank estimator $\widehat{\rho}_k$ from the data convex set $\mathcal{C}_k$ that ultimately either converges to $\rho_t$ for noiseless data, or is very close to it for $N\gg1$. This is akin to conventional CS where for a properly chosen compressive measurement map $\mathcal{M}$, the unique CS estimator, that converges to $\rho_r$ for noiseless data, essentially has the lowest rank out of data convex set. In order to establish a feasible scheme for finding the lowest-rank $\widehat{\rho}_k$ in $\mathcal{C}_k$, we minimize a concave function over $\mathcal{C}_k$ that has rank-minimizing characteristics. For this purpose, two exemplifying concave functions with a global minimum for pure states shall be numerically studied. They are the von Neumann entropy function $S(\rho')=-\tr{\rho'\,\log\rho'}$, and the linear entropy function $S_\textsc{l}(\rho')=1-\tr{\rho'^2}$, both of which are similar in value near all pure states and take the minimum value 0 for the pure states. We mention that entropy minimization was numerically reported in~\cite{Tran:2016aa} to offer a stronger compressive recovery for general low-rank matrices compared to the standard trace-norm minimization in CS.

\begin{figure}[t] 
	\centering
	\includegraphics[width=0.6\columnwidth]{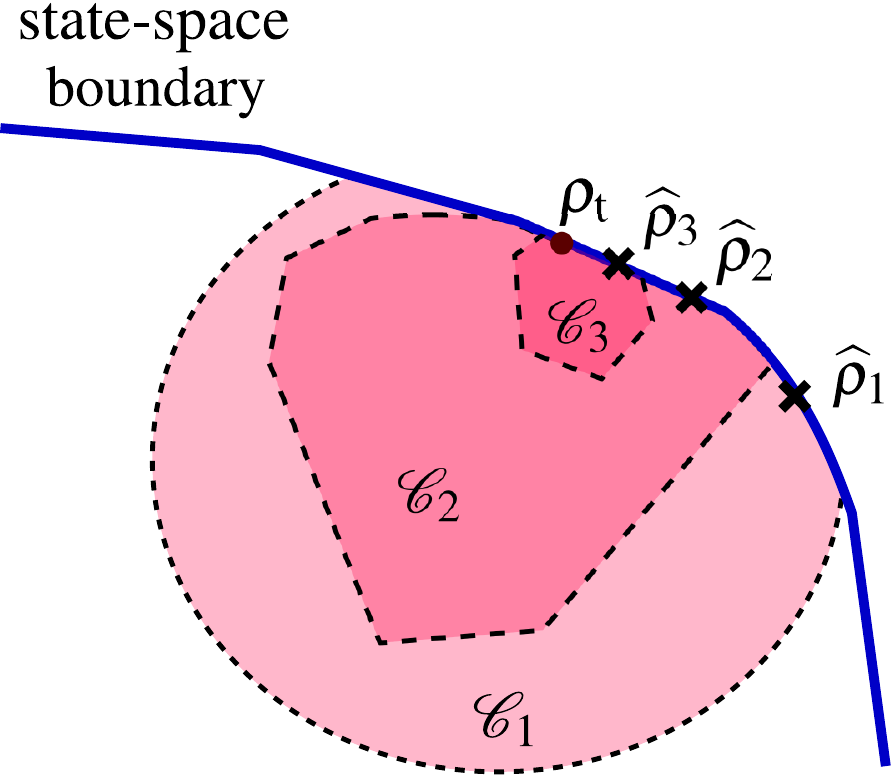}
	\caption{\label{fig:adapt}The progress of the adaptive strategy in ACT. In each step, (linear-)entropy minimization is performed under both positivity and data (ML) constraints to give lowest-(linear-)entropy $\widehat{\rho}_k$. The data convex set $\mathcal{C}_k$ makes a common boundary with the state space. This boundary shrinks as $k$ increases, so that the $\widehat{\rho}_k$s converge to $\rho_r$.}
\end{figure}

While in CS, the measurement map is first decided before a unique lowest-rank estimator is obtained, we note here that for ACT, the mechanism that drives the compressive nature is the interplay between positivity and data (ML) constraints on the entropy minimization procedure, which leads to the correct guidance to the unique $\widehat{\rho}_{k_\textsc{ic}}\approx\rho_r$. It is shown in~\cite{Ahn:2019aa} that owing to these two constraints, if one manages to measure the eigenbasis $\mathcal{B}_{\rho_r}$ of $\rho_r$ that is low-rank ($r\ll d$), then only $k_0=\lceil(r^2-r)/(d-1)\rceil+1\ll d+1$ bases (including $\mathcal{B}_{\rho_r}$) are needed to unambiguously reconstruct $\rho_r$. Therefore the eigenbasis of a lowest-(linear-)entropy $\widehat{\rho}_k$ (which would essentially have an extremely low rank) in $\mathcal{C}_k$ may be assigned to $\mathcal{B}_{k+1}$. In this manner, if $N\gg1$, it is evident that $\mathcal{B}_{k+1}$ approaches $\mathcal{B}_{\rho_r}$, and the action of the optimization procedure under both constraints aid in speeding up this convergence (see Fig.~\ref{fig:adapt}). 

Generally, the adaptive strategy discussed here yields a sequence of entangled bases. For many-body quantum systems, such bases are often difficult to be realized experimentally. To carry out the adaptive strategy for such systems, we impose, in every iterative step, an additional local structure on the adaptive von Neumann bases by searching for the nearest tensor-product basis that is ``closest'' to the optimal entangled one. One way to do so is to first minimize the distance between the lowest-(linear-)entropy $\widehat{\rho}_k\in\mathcal{C}_k$ and another state $\rho'$ with product eigenbasis with respect to some metric, and take the eigenbasis of the optimum. This distance reduction helps to find a local von Neumann basis that is close to $\mathcal{B}_{\widehat{\rho}_k}$ for the lowest-(linear-)entropy $\widehat{\rho}_k$. 

Putting everything together, we arrive at the complete (P)ACT procedure:
\begin{center}
	\begin{minipage}[c][10.5cm][c]{0.9\columnwidth}
		\noindent
		\rule{\columnwidth}{1.5pt}\\
		\textbf{(P)ACT procedure}\\
		\noindent
		Beginning with $k=1$ and a random computational basis $\mathcal{B}_1$:
		\begin{enumerate}
			\item Measure $\mathcal{B}_k$ and collect the relative frequency data $\sum^{d-1}_{j'=0}\nu_{j'k}=1$.
			\item From $\left\{\nu_{0k'},\ldots,\nu_{d-1\,\,k'}\right\}^k_{k'=1}$, obtain $kd$ physical ML probabilities.
			\item Perform ICC with the ML probabilities and compute $s_{\textsc{cvx},k}$:
			\begin{itemize}
				\item \textbf{If}~$s_{\textsc{cvx},k}<\epsilon$, terminate ACT and take $\rho_\text{max}\approx\rho_\text{min}$ as the estimator and report $s_{\textsc{cvx},k}$.
				\item \textbf{Else}~Proceed.
			\end{itemize}
			\item Choose a lowest-(linear-)entropy $\widehat{\rho}_k\in\mathcal{C}_k$ in $\mathcal{C}_k$
			\item Define $\mathcal{B}_{k+1}$ to be the eigenbasis of $\widehat{\rho}_k$ for ACT, or a local basis close to it for PACT \emph{via} some prechosen distance minimization technique.
			\item Set $k=k+1$ and repeat.\\[-5ex]
		\end{enumerate}
		\rule{\columnwidth}{1.5pt}
	\end{minipage}
\end{center}

\section{Simulation Specifications}
\label{sec:sim}

In this brief section, we clarify all essential technical details involved in our subsequent numerical studies. We begin by summarizing the main goals of our numerical studies:
\begin{enumerate}[label={(\Alph*)}]
	\item Compare (P)ACT with the following random measurement schemes~(refer to Sec.~\ref{subsec:csmeas}):
	\begin{enumerate}[label={(\Roman*)}]
		\item Random Pauli~(RP) bases.
		\item Random Haar-uniform~(RH) bases.
		\item Eigenbases of random full-rank states~(RS).
	\end{enumerate}
	\item Benchmark (P)ACT using known analytical scalings of
	\begin{enumerate}[label={(\Roman*)}]
		\setcounter{enumii}{3}
		\item Baldwin--Flammia~(BF) POVMs,
		\item Baldwin--Goyeneche~(BG) bases,
		\item Kech--Wolf scaling~(KW) for arbitrary bases,
	\end{enumerate}
	and conjecture asymptotic $k_\textsc{ic}$ scalings for (P)ACT.
	\item Present and numerically benchmark a new hybrid compressive scheme (HCT).
\end{enumerate}

In what follows, all simulations are performed on ensembles of random quantum states of various ranks $r$ over which important indicators such as $s_\textsc{cvx}$ and $k_\textsc{ic}$ are averaged. To generate these ensembles, we follow the prescriptions in \cite{Zyczkowski:2003hs} and distribute the random states according to the Hilbert-Schmidt measure. For a fixed $r$, this is done by first generating $r\times d$ matrices $A_j$ with entries i.i.d. standard Gaussian distribution, and next define the ensemble density matrices $\{\rho_j\}$ in accordance with $\rho_j=A_j^\dag A_j/\tr{A_j^\dag A_j}$.

Both minimum entropy schemes (with $S$ and $S_\textsc{l}$) over convex sets are carried out efficiently using the superfast accelerated projected gradient algorithm~\cite{Shang:2017sf}. The ICC algorithm is carried out with the CVXOPT package~\cite{cvx,gb08}. While the deterministic compressive measurements, namely the BF POVMs and BG bases, possess analytical $k_\textsc{ic}$ scaling behaviors that can be readily used for benchmarking purposes, the random measurements, namely RP, RH and RS bases have at most approximated scaling expressions. The generations of both RH and RS bases are described in the appendix. For all \emph{local} bases schemes, which refer to PACT, (I), (II), and (III), since the quantum systems that we shall be investigating are multi-qubit, these schemes involve basis outcomes defined as projectors onto the tensor-product space of single-qubits.

For the numerical simulation tasks (a)--(c), the results of which are presented in Sec.~\ref{sec:num_res}, we study all results for noiseless data to understand the underlying theoretical relationships between $\{s_\textsc{cvx},k_\textsc{ic}\}$ and $r$.

\section{Numerical results}
\label{sec:num_res}

\subsection{Comparisons of (product) adaptive compressive tomography with random-bases measurements}

\begin{figure*}[htp] 
	\centering
	\includegraphics[width=1.9\columnwidth]{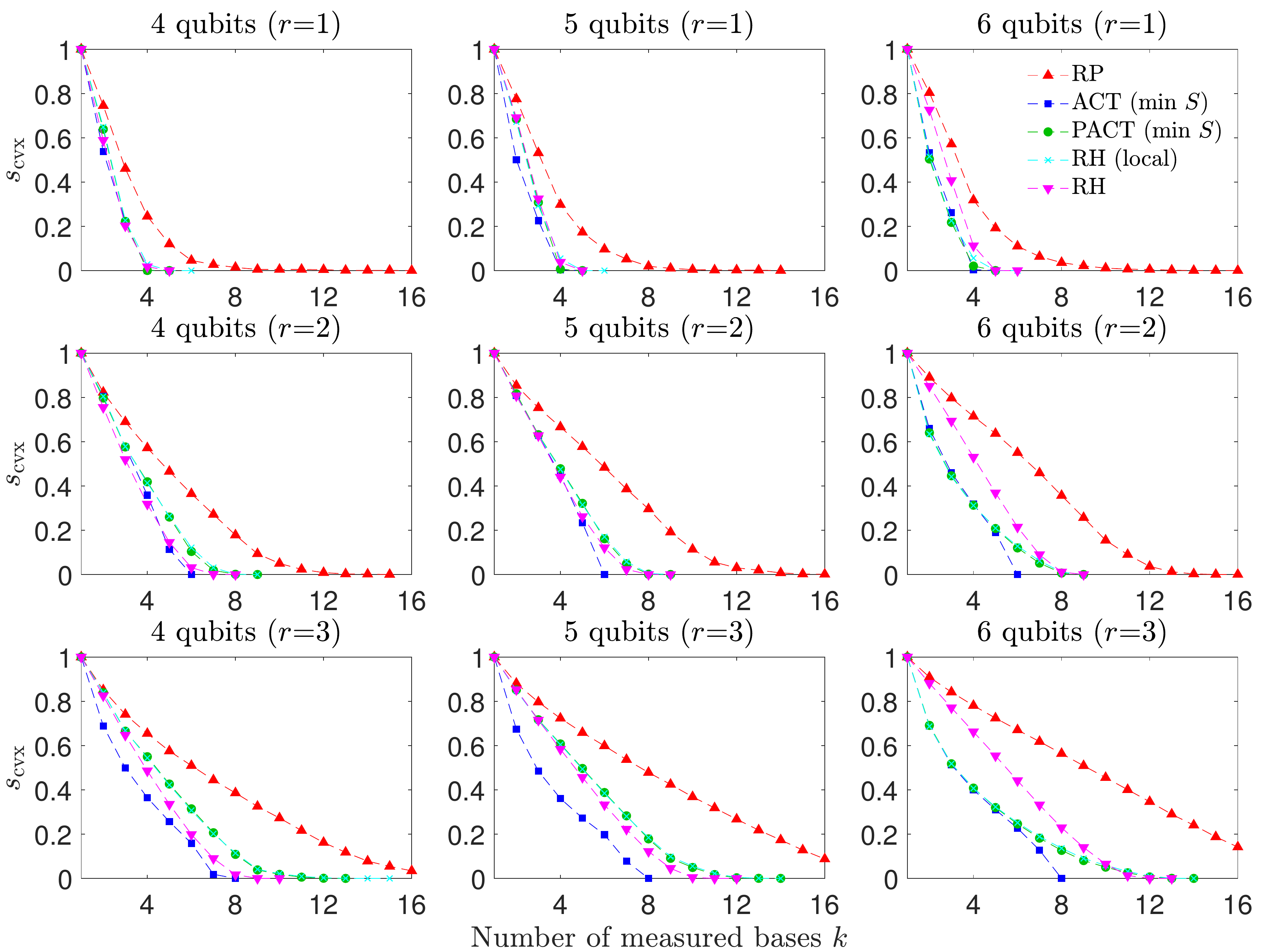}
	\caption{\label{fig:scvx}Plots of $s_\textsc{cvx}$ for multi-qubit systems $d=16$ (4-qubit), 32~(5-qubit) and 64~(6-qubit) and $1\leq r\leq 3$. The data markers for each system and $r$ is averaged over 100 random states that are uniformly distributed with respect to the HS measure.}
\end{figure*}

\begin{figure*}[htp] 
	\centering
	\includegraphics[width=2\columnwidth]{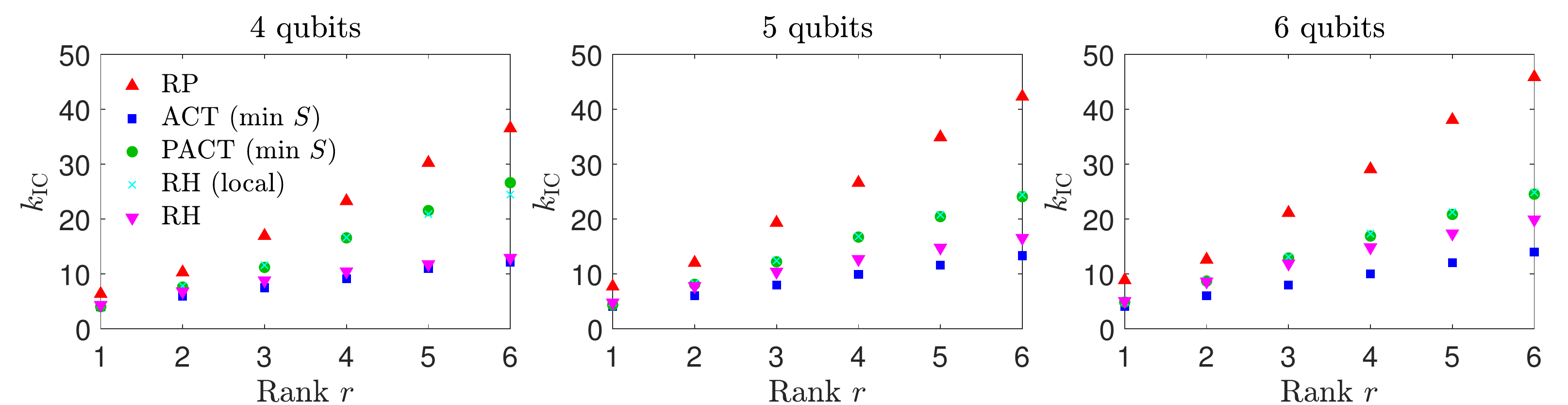}
	\caption{\label{fig:kic}Plots of simulation of $k_\textsc{ic}$, which shows a different aspect of the ICC performance for all schemes than Fig.~\ref{fig:scvx}.}
\end{figure*} 

\begin{figure*}[htp]
	\centering
	\includegraphics[width=1.9\columnwidth]{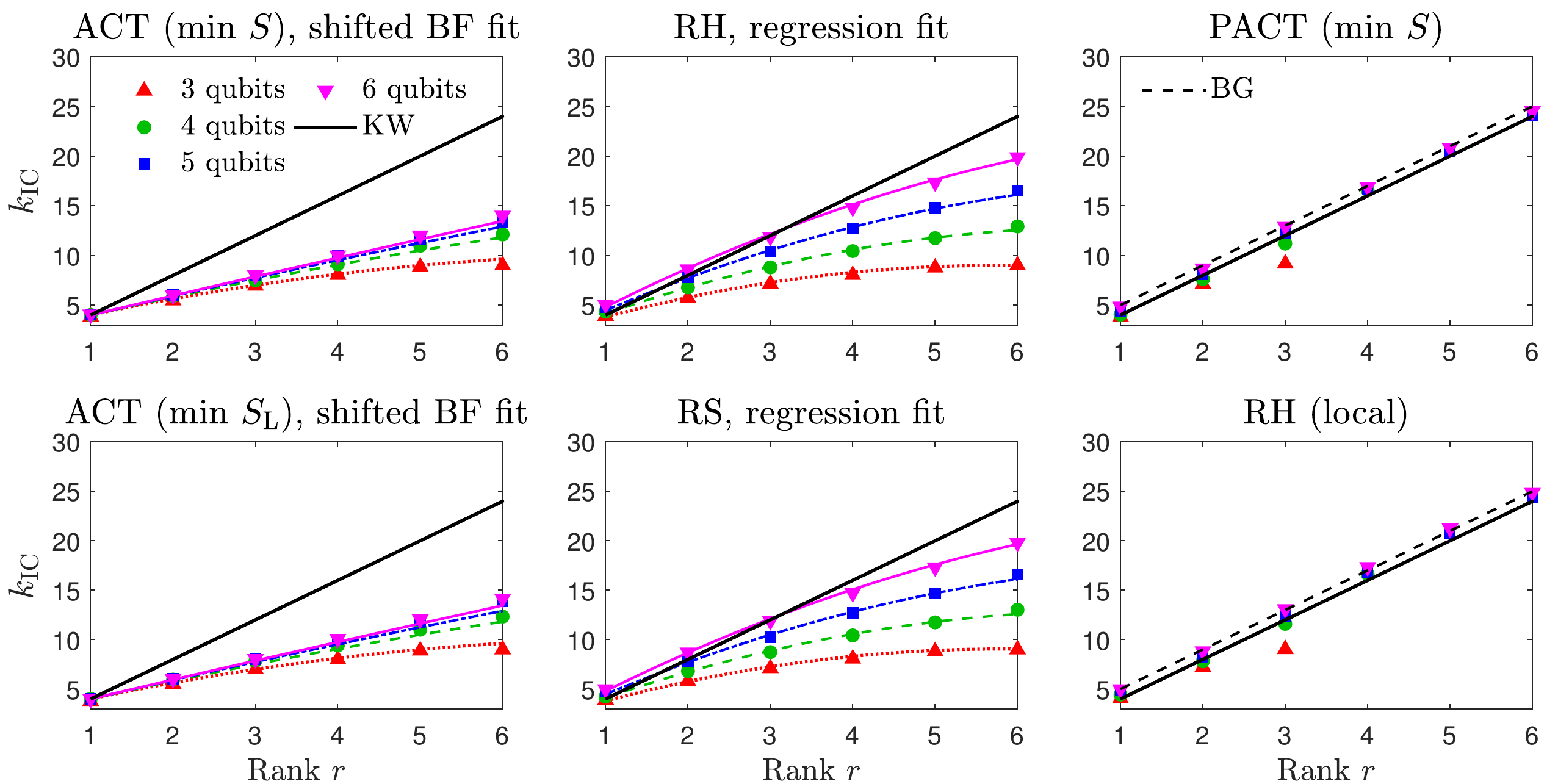}
	\caption{\label{fig:kic_scaling}A comparison of various schemes for different multi-qubit systems and some of their behaviors relative to BF POVMs and BG bases. Both ACT and PACT follow closely to the respective scaling behaviors of the (shifted) BF POVMs and BG bases. The nonlinear regression fits for RH and RS follow a general $O(r\log d)$ slope with quadratic $r$-dependent intercepts that are relatively weak in the limit $r\ll d$. The linear KW benchmark overestimates both ACT schemes for $r>1$, and does so as well for RH and RS schemes. Its linear behavior lies close to that of the BG bases scheme, with the latter having very similar scaling as the asymptotic behaviors for both PACT and RH bases scheme.}
\end{figure*}

We proceed to systematically compare both ACT and PACT respectively with popular random compressive schemes (I)--(III) in Sec.~\ref{sec:sim}. Figure~\ref{fig:scvx} shows the results of ICC for true quantum states of 4-, 5- and 6-qubit systems. For each type of system, ICC is applied to all schemes on states of ranks $r=2$, 4 and 6. For all the schemes, we confirm that $s_{\textsc{cvx},k}$ monotonically decreases over the number of measured von Neumann bases $k$, and reach $s_{\textsc{cvx},k_\textsc{ic}}=0$ (up to some numerical threshold). In terms of the convergence rate of $s_{\textsc{cvx},k}$, it turns out that (P)ACT as well as schemes (II)--(V) are more efficient than RP. Here, ACT scheme shows most efficient performance. Faster convergence of $s_{\textsc{cvx},k}$ directly leads to smaller $k_\textsc{ic}$, which implies higher compression in size of IC data. 

The compression efficiencies for all schemes are compared \emph{via} $k_\textsc{ic}$ more clearly in Fig.~\ref{fig:kic}, which validates that ACT is the most efficient scheme for all number of qubits, whereas RP turns out to be the most inefficient one. More specifically, it is evident that the $k_\textsc{ic}$ gap between ACT and RH, and that between PACT/RH~(local) and RP enlarges for larger $r$ and number of qubits, which implies that ACT is relatively more for states of higher rank and Hilbert-space dimension. The apparent coincidence between PACT and RH~(local) for all $d$ may be attributed to an unavoidable minimum level of incurred randomness in PACT that arises from the restriction to local-bases measurement during adaptive optimization. Such a level of randomness is sufficient to practically render PACT equivalent to RH.

\subsection{Benchmarks for (product) adaptive compressive tomography and asymptotic performances}

Benchmarking of the $k_\textsc{ic}$ performance for ACT, RH, RS, PACT and local RH schemes is done with the known analytical standards (IV)--(VI) listed in Sec.~\ref{sec:sim}, and the results are presented in Fig.~\ref{fig:kic_scaling}. The ACT schemes carried out through minimizing $S$ and $S_\textsc{l}$ are compared with the element-probing BF POVMs for arbitrary rank-$r$ states, which contains outcomes of non-unit rank and possess a total number of $M=(2dr-r^2+1)$ outcomes. The figure shows that for both ACT protocols, $k_\textsc{ic}$ scaling behaviors are in good agreement with the expression $M/d+2$ (a shifted BF scaling). This means that ACT, which yields only rank-1 von Neumann projectors, exceeds only by 2 bases in performance as compared to BF POVMs to perform IC state tomography.

We also note that both ACT schemes defined by the minimization of two different entropy functions give almost indistinguishable $k_\textsc{ic}$ curves for all tested $r$ and $d$. This numerical observation confirms that the behaviors of von Neumann and linear entropies are almost the same with respect to the optimization algorithm in the adaptive stage. Thus, the (P)ACT protocols involving the von~Neumann entropy minimization is sufficient for the remaining discussions. 
Indistinguishable $k_\textsc{ic}$ curves exist also for random schemes (such as RH and RS), and this amply hints that the effect of informational completeness for the non-adaptive schemes depends weakly on the specific choice of random bases generation algorithm, but rather more strongly on the state rank.  

\begin{figure}[h!]
	\centering
	\includegraphics[width=0.85\columnwidth]{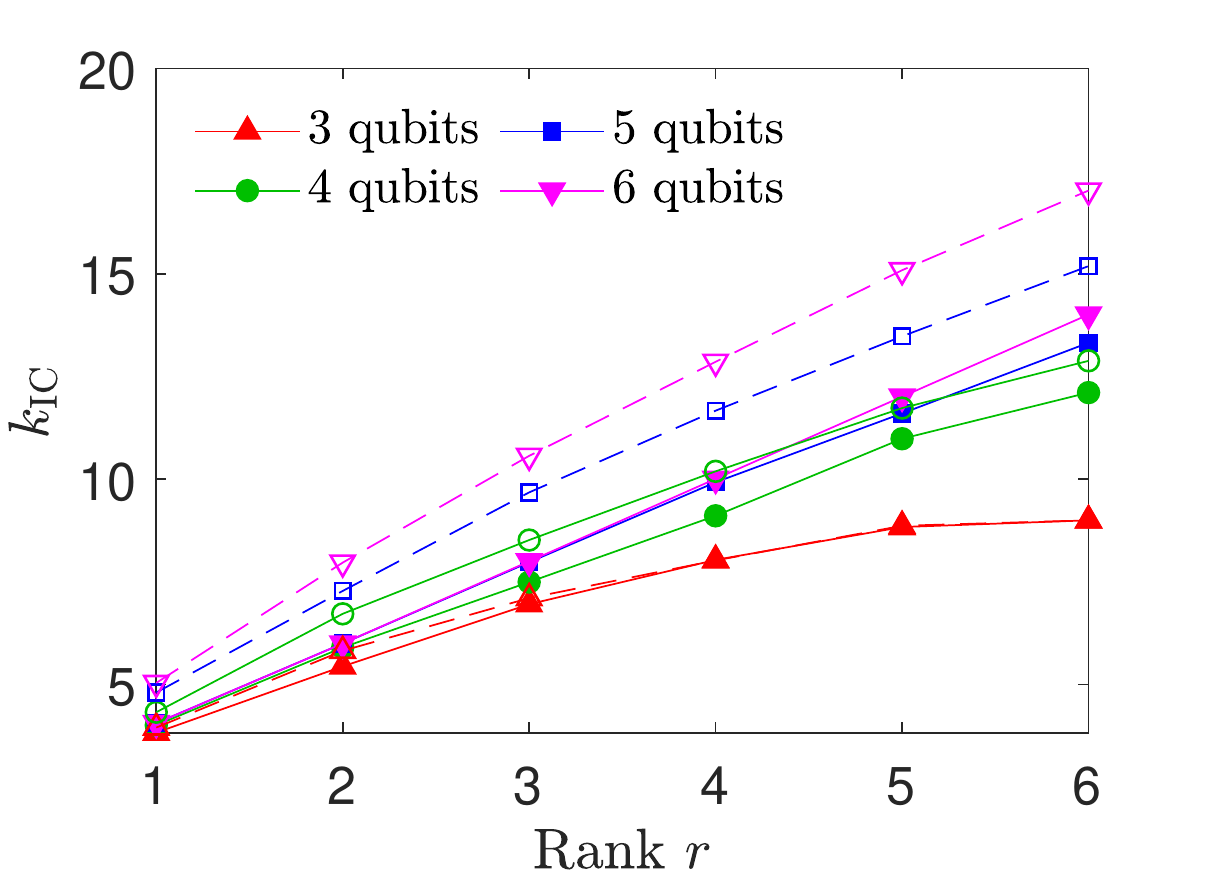}
	\caption{\label{fig:ACT_vs_rdfc}Plots of $k_\textsc{ic}$ for min-$S$ ACT (filled markers and solid curves) and the strategy of picking eigenbases of random rank-deficient states from data convex sets (unfilled markers and dashed curves). The discrepancies are magnified for large $d$.}
\end{figure}

One might, at this point, naively suppose that perhaps ACT should also work sufficiently well if $\widehat{\rho}_k$ is some randomly chosen rank-deficient state in $\mathcal{C}_k$ for every $k$. Fig.~\ref{fig:ACT_vs_rdfc} compares the average $k_\textsc{ic}$ of min-$S$ adaptive bases with that of eigenbases of random rank-deficient states. It presents clear numerical evidence contrary to the above supposition, which further reinforces the statement that unlike random schemes, an appropriate choice of objective function is necessary for ACT to achieve optimal compression.

In the context of many-body local compressive measurements, we compare the $k_\textsc{ic}$s of PACT and local RH scheme with those of the BG bases for arbitrary $r$ and $d$, the latter which employs $4r+1$ specifically constructed orthonormal entangled bases. Figure~\ref{fig:kic_scaling} tells us that both product schemes asymptotically approach the BG scheme in $k_\textsc{ic}$ performance as the number of qubits grows.
 
\begin{figure}[t]
	\centering
	\includegraphics[width=1\columnwidth]{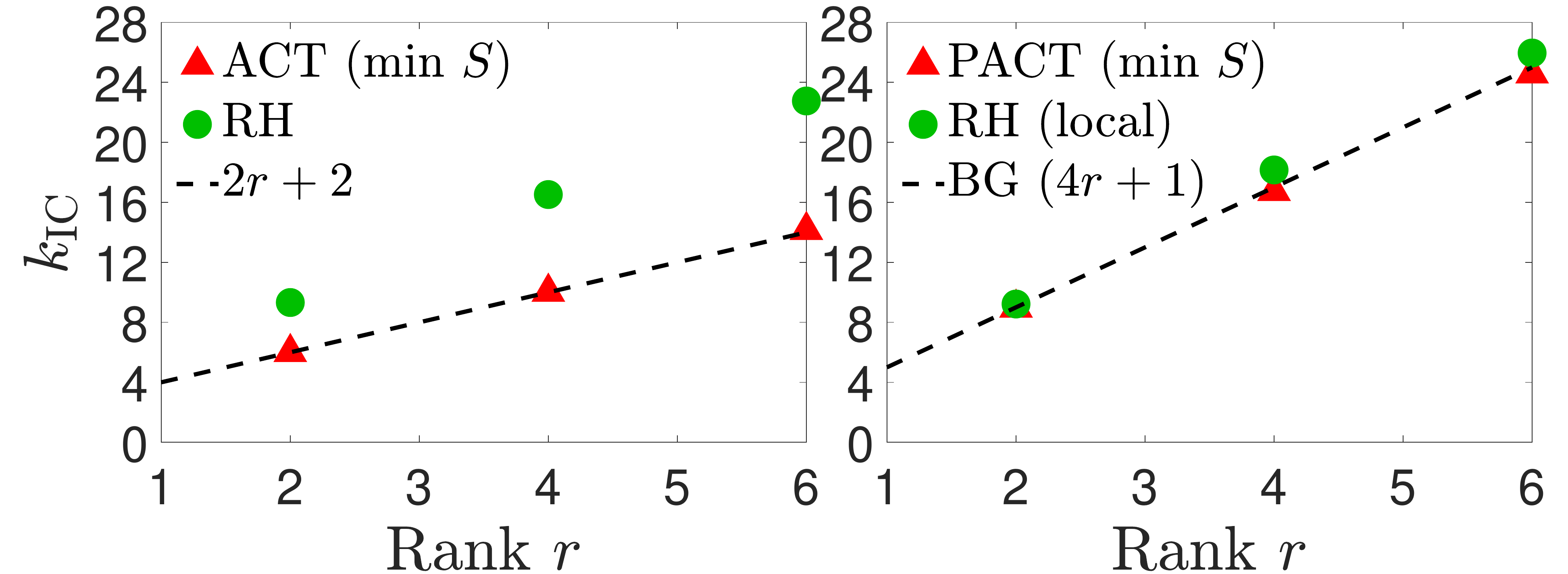}
	\caption{\label{fig:kic_scaling_d128}Plots of $k_\textsc{ic}$ scaling behaviors for 7-qubit systems ($d=128$), averaged over 50 random states per data marker. The scaling conjectures $2r+2$ and $4r+1$ still stand accurately for the sampled $r$ values.}
\end{figure}

Hence, we numerically confirm that for all $r$ and $d$, ACT shows stronger compressive behavior than PACT and non-adaptive random schemes. The result of benchmarking with BF POVMs and BG bases reveals $k_\textsc{ic}$ behaviors for ACT and PACT that lose the dependency on $d$ in limit of large number of qubits. These conjectured asymptotic scalings are respectively $k_\textsc{ic}=2r+2$ and $k_\textsc{ic}=4r+1$, and our belief in their validity is further strengthened with 7-qubit systems in Fig.~\ref{fig:kic_scaling_d128}.

All the compressive bases schemes presented here are contrasted with the KW benchmark that applies to arbitrary bases measurements, which is effectively a linear function of $r$ that is independent of $d$ due to the ceiling function. We point out that the KW benchmark that estimates the required number of bases needed to perform IC reconstruction using arbitrary von Neumann bases almost always overestimates the $k_\textsc{ic}$ performance of ACT. The reason is that the measurement resources for ACT scales nonlinearly with $r$ owing to the low-rank guidance from both positivity and data constraints. The KW benchmark also show signs of overestimation for the random RH and RS schemes for sufficiently large $r$.

\subsection{Hybrid compressive tomography}
\label{sec:hybrid}

While the complete ACT scheme is highly compressive ($k_\textsc{ic}=2r+2$), the computational resources needed to search for the optimal estimators in the data convex set will eventually become noticeably expensive for extremely large physical systems. On the other hand, a fully random scheme simply suggests measurement bases randomly and requires essentially negligible computational resources. 

As an attempt to speed up the compression process, we may take advantage of the benefits from both random and adaptive protocols and establish a hybrid compressive tomography~(HCT) scheme. This hybrid scheme starts with choosing random von Neumann measurement bases, which may be constructed with either the RH or RS prescription. In fact, any sort of bases would deliver similar performance due to the limited data one always has at the beginning of a compression process. As the accumulated dataset is built up, more information is gained about the unknown $\rho_r$ such that it is now justified to spend more effort in searching for good optimal measurements based on more statistically reliable data. Armed with this basic insight, we propose the following procedure:

\begin{center}
	\begin{minipage}[c][12cm][c]{0.9\columnwidth}
		\noindent
		\rule{\columnwidth}{1.5pt}\\
		\textbf{HCT procedure}\\
		\noindent
		Beginning with $k=1$, a random computational basis $\mathcal{B}_1$ and some positive threshold value $s_{\mathrm{th}}$:
		\begin{enumerate}
			\item Measure $\mathcal{B}_k$ and collect the relative frequency data $\sum^{d-1}_{j'=0}\nu_{j'k}=1$.
			\item From $\left\{\nu_{0k'},\ldots,\nu_{d-1\,\,k'}\right\}^k_{k'=1}$, obtain $kd$ physical ML probabilities.
			\item Perform ICC with the ML probabilities and compute $s_{\textsc{cvx},k}$:
			\begin{itemize}
				\item \textbf{If}~$s_{\textsc{cvx},k}<\epsilon$, terminate ACT and take $\rho_\text{max}\approx\rho_\text{min}$ as the estimator and report $s_{\textsc{cvx},k}$.
				\item \textbf{Else}~Proceed.
			\end{itemize}
			\item If $s_{\textsc{cvx},k}>s_{\mathrm{th}}$:
			\begin{itemize}
				\item Assign a random basis, perhaps from Appendix~\ref{app:rand}, to $\mathcal{B}_{k+1}$.
			\end{itemize}
			\noindent
			Else:
			\begin{itemize}
				\item Assign an optimal basis obtained using the adaptive strategy in (P)ACT to $\mathcal{B}_{k+1}$.
			\end{itemize}
			\item Set $k=k+1$ and repeat.\\[-5ex]
		\end{enumerate}
		\rule{\columnwidth}{1.5pt}
	\end{minipage}
\end{center}

\begin{figure}[t]
	\centering
	\includegraphics[width=1\columnwidth]{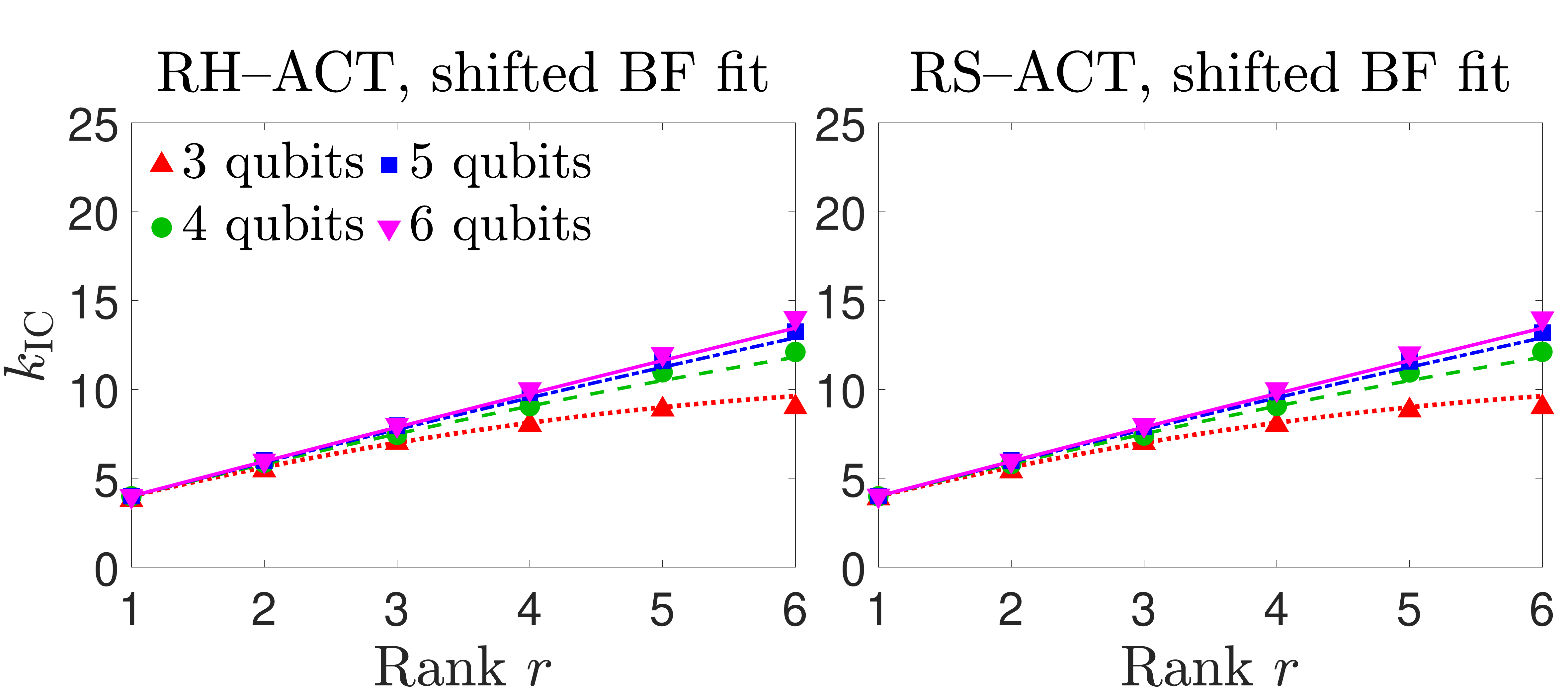}
	\caption{\label{fig:kic_scaling_hybrid}Plots of $k_\textsc{ic}$ for HCT with $s_\text{th}=0.5$. Both sets of graphs match almost exactly with the two ACT plots in Fig.~\ref{fig:kic_scaling}.}
\end{figure}

\begin{figure*}[t]
	\centering
	\includegraphics[width=0.9\textwidth]{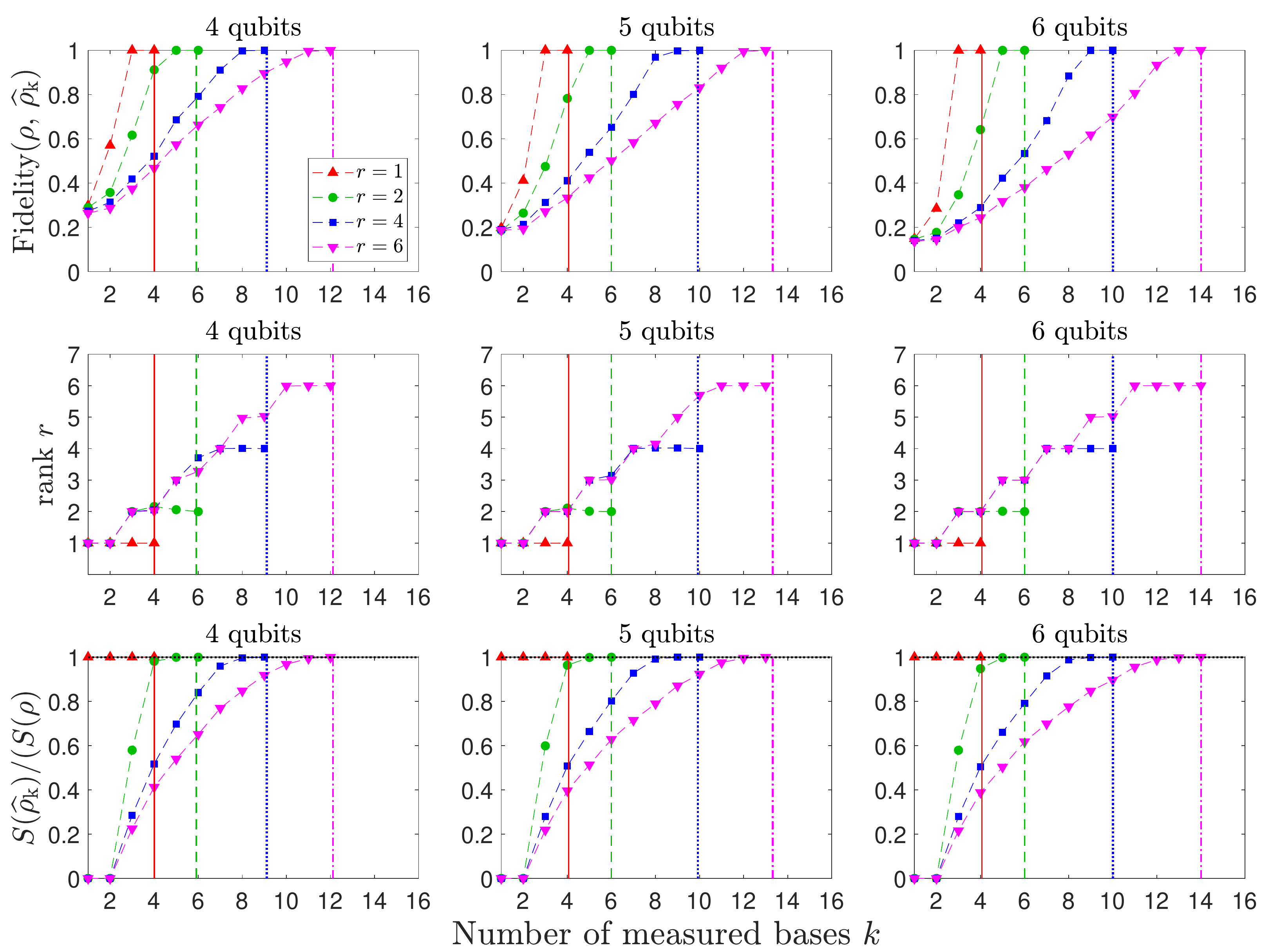}
	\caption{\label{fig:qualities}Noiseless evolution of the different minimum-entropy estimator qualities as ACT progresses. Vertical lines mark the $k_\textsc{ic}$ values for the respective ranks $r$. All curves are averaged over 100 random states that are uniform in the HS measure per $r$ value. Here, $S(\rho)$ is only a convenient normalization factor for averaging $S(\widehat{\rho}_k)$ over true states.}
\end{figure*}

Figure~\ref{fig:kic_scaling_hybrid} reveals that both RH--ACT and RS--ACT type HCT schemes give almost the same averaged $k_\textsc{ic}$ behaviors, both of which are essentially identical to ACT as in Fig.~\ref{fig:kic_scaling}. Based on these numerical evidences, we may argue that the cause of these equivalent performances comes from almost identical gradual average gains in tomographic information throughout the respective compressive tomography schemes. This supports our observation that, without relying on any \emph{a priori} knowledge about $\rho_r$, adaptive methods only serve as crucial roles in compressive tomography after sufficient amount of tomographic information is acquired for these methods to yield more correct optimized measurement bases. Otherwise, the identical average compression capabilities of both ACT and HCT imply that both schemes effectively acquire the same amount of tomographic information at the early stages of the processes. This justifies the more economical approach of first measuring random bases during the initial stage of HCT, which is far less time-consuming than optimizing for adaptive bases in ACT on highly complex quantum systems.

\section{Epilogue: Remarks on informational completeness in adaptive compressive tomography}
\label{sec:epilogue}

Besides $s_\textsc{cvx}$, we may inspect other aspects of the data convex set in ACT, carried out with $S$ minimization for example, to preempt its correct termination in case one intends to stop the reconstruction procedure with $k<k_\textsc{ic}$ measurement bases. Figure~\ref{fig:qualities} showcases the average behaviors of various different minimum-entropy $\widehat{\rho}_k$ qualities for ACT under negligible statistical fluctuation ($N\gg1$). We see that the fidelity, rank and entropy of $\widehat{\rho}_k$ reach the correct true values as $k$ increases, and saturate at $k<k_\textsc{ic}$. Technically, both the fidelity and rank cannot be used to judge the termination of ACT, since the fidelity requires knowledge of $\rho_r$, and it is also impossible to guess whether the rank of $\widehat{\rho}_k$ is close to the correct value without $\mathrm{rank}\{\rho_r\}$ as reference because of its regular stepped gradient. However, based on our numerical experience, the monotonically increasing entropy $S(\widehat{\rho}_k)$ (a direct mathematical consequence of the inequality chain $\mathcal{C}_1\supseteq\mathcal{C}_2\supseteq\ldots\supseteq\mathcal{C}_{k_\textsc{ic}}$ for noiseless data) always approaches the final true value with smoothly decreasing gradient $S(\widehat{\rho}_{k})-S({\widehat{\rho}_{k-1}})<S(\widehat{\rho}_{k-1})-S(\widehat{\rho}_{k-2})$. We may then attempt to prematurely stop ACT at $k=k_0$ when $S(\widehat{\rho}_{k_0})-S(\widehat{\rho}_{k_0-1})$ is small enough and use the resulting estimator $\widehat{\rho}_{k_0}$ as an \emph{approximately IC reconstruction} for future statistical predictions, in which case this low-rank $\widehat{\rho}_{k_0}$ will be close to $\rho_r$ even though the size of $\mathcal{C}_{k_0}$ can be appreciably greater than zero. With statistical consistency, this termination method works also for real data of sufficiently large $N$ (or low statistical fluctuation).

The aforementioned remarks eventually lead to a subtle point behind the notion of ``IC'' adopted by ACT. In general tomography settings, which include the studies in CS, one typically defines ``IC'' in terms of the class of quantum states of interest, such as the class of rank-$r$ states to name one. In ACT, however, we speak of an IC dataset $\mathbb{D}$ that is statistically related to the particular unknown state we are probing. As a result, the notion of ``IC'' in ACT, along with the value of $k_\textsc{ic}$, strongly depends on both the adaptive bases and $\mathbb{D}$.

\begin{figure}[t] 
	\centering
	\includegraphics[width=1\columnwidth]{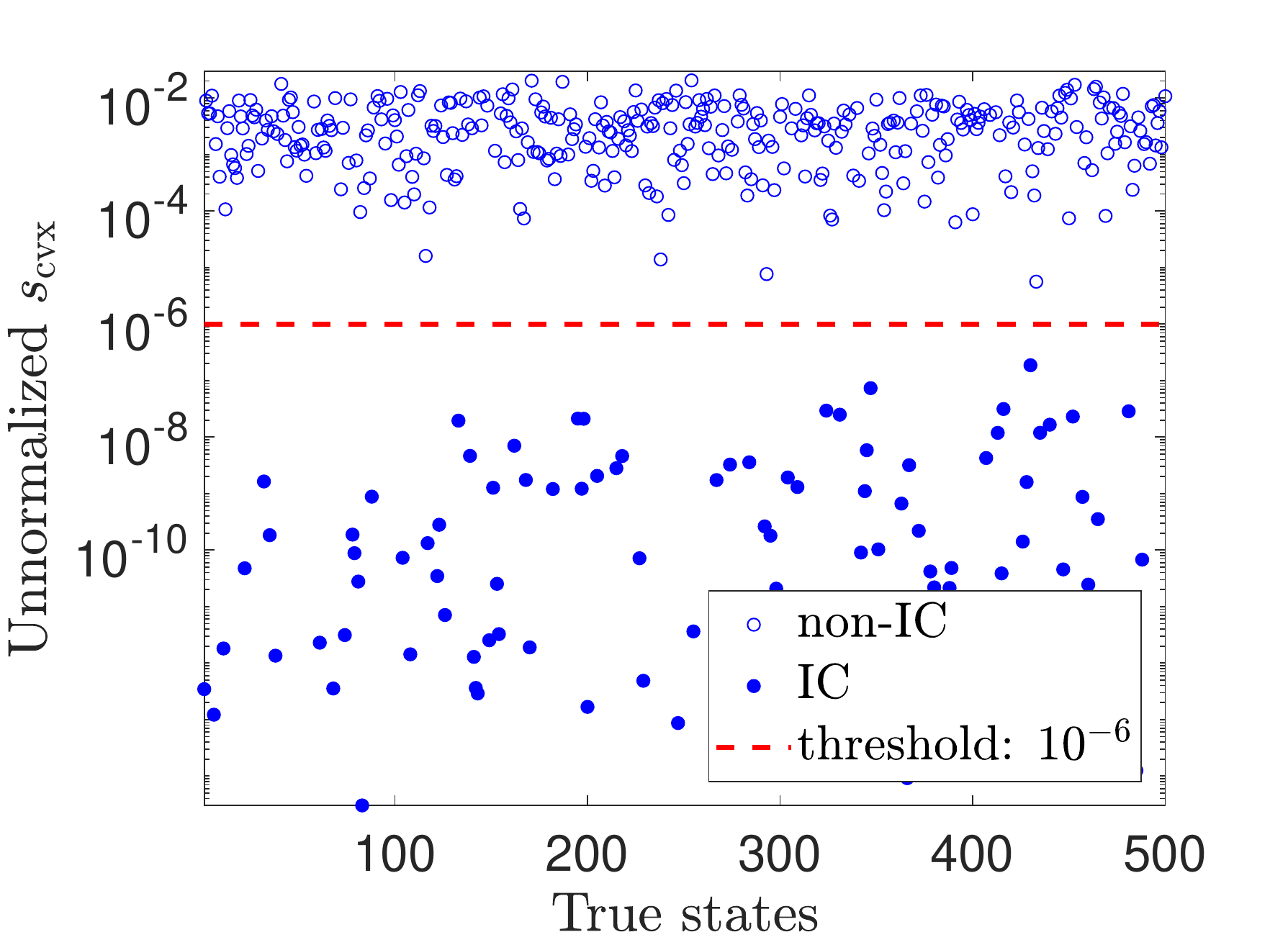}
	\caption{\label{fig:IC}Noiseless distribution of $s_\textsc{cvx}$ for a Haar-uniform set of 500 rank-one 5-qubit $\rho_r$s ($d=2^5=32$) based on a fixed set of ACT bases that is IC for another pure state. The decision for labeling the probabilities as ``IC'' typically only weakly relies on the choice of a threshold so long as it is reasonable. Since truly IC situations correspond to very small $s_\textsc{cvx}$ close to the numerical limits of the semidefinite program in ICC, which are clearly separated from IC situations of significantly larger $s_\textsc{cvx}$ as shown in this figure, a reasonable threshold value below which all points are IC may be conveniently fixed at $10^{-6}$.}
\end{figure}

A more objective $\mathbb{D}$- or $\rho_r$-dependent compressive strategy in exchange for spurious or ad hoc prior assumptions about $\rho_r$ also means that an ACT bases POVM that is IC for a given unknown rank-$r$ state is not necessarily going to be IC for another rank-$r$ state. Figure~\ref{fig:IC} elucidates this fact for a distribution of rank-one true states probed by an ACT bases POVM that is IC for a particular pure state assuming noiseless probabilities.
  
\section{Conclusions}

We have performed a comprehensive numerical study of our adaptive compressive tomography schemes, which can uniquely reconstruct any arbitrary rank-deficient quantum state with few measurement bases than the square of its given dimension without any other \emph{a priori} information about this state. Several numerical results in this article can serve as important guidelines for resource planning, structuring and executing objective compressive tomography for multi-qubit systems, which are widely used in quantum information and quantum optics studies.

After simulating up to reasonably large dimensions $d$ of multi-qubit systems and quantum-state rank $r$, we can summarize these numerical results under low statistical fluctuation:
\begin{enumerate}
	\item The average performances (minimal number of informationally complete bases $k_\textsc{ic}$) of adaptive compressive tomography schemes almost always beat those of random compressive schemes, which include random Pauli bases, random Haar-unitary bases and eigenbases of random Hilbert-Schmidt-uniform states.
	\item The average $M_\textsc{ic}=k_\textsc{ic}d$ for the entangled adaptive compressive scheme is greater than that of Baldwin-Flammia measurement outcomes by only $2d$ for all tested range of $r$ and $d$. The asymptotic ($d\gg2$) average $k_\textsc{ic}$ for both entangled and product adaptive schemes are respectively $2r+2$ and $4r+1$.
	\item There is virtually no difference in average $k_\textsc{ic}$ performance between a fully adaptive scheme and a hybrid scheme that starts off as a random scheme followed by an adaptive scheme at a later stage for a reasonable transition point. Therefore, the hybrid compressive scheme may be used to speed up the tomography compression process.
\end{enumerate}

\section*{acknowledgments}
We acknowledge financial support from the BK21 Plus Program (21A20131111123) funded by the Ministry of Education (MOE, Korea) and National Research Foundation of Korea (NRF), the NRF grant funded by the Korea government (MSIP) (Grant No. 2010-0018295), the European Research Council (Advanced Grant PACART), the Spanish MINECO (Grant FIS2015-67963-P), the Grant Agency of
the Czech Republic (Grant No. 18-04291S), and the IGA Project of the
Palack{\'y} University (Grant No. IGA PrF 2019-007).

\appendix

\section{List of acronyms}
\label{app:acro}

\begin{tabular}{rl}
	ACT: & \quad adaptive compressive tomography\\
	BF: & \quad Baldwin--Flammia\\
	BG: & \quad Baldwin--Goyeneche\\
	CS: & \quad compressed sensing\\
	HCT: & \quad hybrid compressive tomography\\
	IC: & \quad informationally complete\\
	ICC: & \quad informational completeness certification\\
	KW: & \quad Kech--Wolf\\
	ML: & \quad maximum-likelihood\\
	PACT: & \quad product adaptive compressive tomography\\
	POVM: & \quad positive operator-valued measure\\
	RH: & \quad random Haar\\
	RP: & \quad random Pauli\\
	RS: & \quad eigenbasis of random states
\end{tabular}

\section{Constructions of random bases}
\label{app:rand}

We briefly supply two short routines to construct von Neumann bases that are generated by random Haar-distributed unitary operators~(RH bases), as well as those that are eigenbases of random full-rank states uniformly distributed according to the Hilbert-Schmidt measure~(RS bases). These two unitary sets have generally different operator probability distributions, which may be verified by comparing some of their operator moments.

It is well-known that the QR decomposition generates unitary operators distributed according to the Haar measure~\cite{Mezzadri:2007qr}, so that the following routine applies:
\begin{center}
	\begin{minipage}[c][8cm][c]{0.9\columnwidth}
		\noindent
		\rule{\columnwidth}{1.5pt}\\
		\textbf{Constructing an RH basis}\\[1ex]
		Starting from a reference basis $\mathcal{B}_\text{ref}=\{\ket{0},\ket{1},\ldots,\ket{d-1}\}$:
		\begin{enumerate}
			\item Generate a random $d\times d$ matrix $A$ with entries i.i.d. standard Gaussian distribution.
			\item Compute $Q$ and $R$ from the QR decomposition $A=QR$.
			\item Define $R_\text{diag}=\mathrm{diag}\{R\}$.
			\item Define $L=R_\text{diag} \oslash |R_\text{diag}|$ ($\oslash$ refers to the Hadamard division).
			\item Define $U_\mathrm{Haar}=QL$.
		\end{enumerate}
		Thereafter construct the new basis $\mathcal{B}_\mathrm{Haar}=U_\mathrm{Haar}\mathcal{B}_\text{ref}$.
		
		\rule{\columnwidth}{1.5pt}
	\end{minipage}
\end{center}  
Steps 3--5 enforces a QR decomposition procedure that produces an effective ``$R$'' matrix that has positive diagonal entries, which is the correct decomposition procedure we need to generate $U_\mathrm{Haar}$.

The second class may be easily generated by the following pseudocode:
\begin{center}
	\begin{minipage}[c][8cm][c]{0.9\columnwidth}
		\noindent
		\rule{\columnwidth}{1.5pt}\\
		\textbf{Constructing an RS basis}\\[1ex]
		Starting from a reference basis $\mathcal{B}_\text{ref}=\{\ket{0},\ket{1},\ldots,\ket{d-1}\}$:
		\begin{enumerate}
			\item Generate a random $d\times d$ matrix $A$ with entries i.i.d. standard Gaussian distribution.
			\item Define $\rho_\textsc{hs}=\dfrac{A^\dagger A}{\tr{A^\dagger A}}$.
			\item Obtain $U_\textsc{hs}$ from the spectral decomposition $\rho_\textsc{hs}=U_\textsc{hs}D_\textsc{hs}U_\textsc{hs}^\dagger$ of diagonal $D_\textsc{hs}$.
		\end{enumerate}
		Thereafter construct the new basis $\mathcal{B}_\textsc{hs}=U_\textsc{hs}\mathcal{B}_\text{ref}$.
		
		\rule{\columnwidth}{1.5pt}
	\end{minipage}
\end{center}  

\vspace{3cm}


%\bibliography{Biblio}

\end{document}